%% file: main.tex
\documentclass[
    pra,
    superscriptaddress,
    twocolumn,
    aps,
    10pt,
    nofootinbib,
]{revtex4-2}

\usepackage{graphicx}
\usepackage{hyperref}
\usepackage{mathtools}
\usepackage{amsfonts}
\usepackage{xcolor}
\usepackage[en-US]{datetime2}  

\begin{document}

\title{Clock Synchronization with Weakly Correlated Photons}

\author{Justin~Yu~Xiang~\surname{Peh}}
\author{Darren~Ming~Zhi~\surname{Koh}}
\author{Zifang~\surname{Xu}}
\author{Xi~Jie~\surname{Yeo}}
\author{Peng~Kian~\surname{Tan}}
\affiliation{Centre for Quantum Technologies, 3 Science Drive 2, Singapore 117543}
\author{Christian~\surname{Kurtsiefer}}
\email{christian.kurtsiefer@gmail.com}
\affiliation{Centre for Quantum Technologies, 3 Science Drive 2, Singapore 117543}
\affiliation{Department of Physics, National University of Singapore, 2 Science Drive 3, Singapore 117542}

\date{\today}

\begin{abstract}
Clock synchronization is necessary for communication and distributed computing tasks.
Previous schemes based on photon timing correlations use pulsed light or photon pairs for their strong timing correlations.
In this work, we demonstrate successful synchronization of crystal clocks using weakly time-correlated photons of 180\,ns coherence time from a bunched light source.
A synchronization timing jitter of 10\,ns is achieved over symmetric -102\,dB
optical channel loss between two parties, over a span of 25\,hours.
We also present a model that gives better estimates to the coherence peak finding success probabilities under low signal.
\end{abstract}

\maketitle
\section{Introduction}
Clock synchronization is used in everyday tasks such as navigation and distributed computing. This is commonly implemented using the Network Time Protocol or global navigation satellite system (e.g., GPS) time synchronization, achieving precision of milliseconds or tens of nanoseconds, respectively~\cite{mills2006computer,renfro2022analysis}. Quantum communication protocols also require clock synchronization, but on the order of nano- to pico-seconds, and are typically achieved either using pulsed laser light~\cite{giorgetta2013optical} or with classical signals on dedicated optical or electronic channels~\cite{gerritsWhiteRabbitassistedQuantum2022,alshowkanSynchronizingQuantumLocal2022}.

Modern quantum communication systems can use the resources present in the protocol itself to perform the clock synchronization, such as photons
arriving at fixed timing intervals, or photon pairs from spontaneous parametric down-conversion (SPDC)~\cite{lee2019symmetrical,ho2009clock}. This has been demonstrated with frequency standards such as Rubidium (Rb) clocks, which provide a long-term frequency stability of \textless1\,ppb/day. In comparison, crystal oscillators without temperature stabilization have frequency stability of only 100\,ppb/day.

Recent works have shown that Rb clocks are unnecessary and crystal oscillators are sufficient, by relying on the strong timing correlation ($\sim$ps) of SPDC photon pairs~\cite{spiessClockSynchronizationCorrelated2023}. Weak coherent photon pulses were also proposed for clock frequency transfer~\cite{spiessClockSynchronizationPulsed2023}. In both cases, the cross-correlation peak is very strong, i.e., $g^{(2)}(\tau=0) \gg 1$, with resolution generally limited only by the timing jitter of the generation and detection optoelectronics.

Timing correlations also exist in thermal light, which has been used in applications such as ghost imaging~\cite{bennink2002two} and range finding~\cite{tan2023practical}.
These timing correlations arise from temporal photon bunching, also known as the Hanbury-Brown-Twiss effect~\cite{brown1956correlation}.
In particular, the use of thermal light opens up the potential for distributed clock synchronization due to photon bunching being preserved across arbitrary partitioning, and makes it hard to manipulate the timing information due to the randomness of the individual photon timing.

However, unlike photon pairs, identification of this photon bunching peak in the
correlation $g^{(2)}(\tau)$ is more challenging due to the low signal (i.e., $g^{(2)}(\tau)\le{}2$) as well as fluctuations about the background of $g^{(2)} = 1$.
An earlier proposal suggests the use of thermal light for clock synchronization by relying on the use of low efficiency two-photon absorption in single-photon detectors to resolve the bunching characteristic~\cite{zhu2013new,xu2020determination}; such a scheme has yet to be demonstrated.

Here, we demonstrate clock synchronization using one such a weakly timing-correlated light source with $g^{(2)}(\tau=0)=1.42$ and coherence time
$\tau_c = 180$\,ns, over two transmission channels with a loss of
\mbox{-102}\,dB each. We reach an accuracy of 10\,ns, and continuously track the frequency drift over a period of 25\,hours to demonstrate its stability.
Our scheme only requires single-photon detection and crystal oscillators as reference clocks.
We additionally derive the probability of achieving clock synchronization over transmission channels, which is applicable across different time-correlated photon sources, including SPDC pair sources and thermal light sources.

\section{Clock synchronization with bunched light}

\begin{figure*}
  \includegraphics[width=\textwidth]{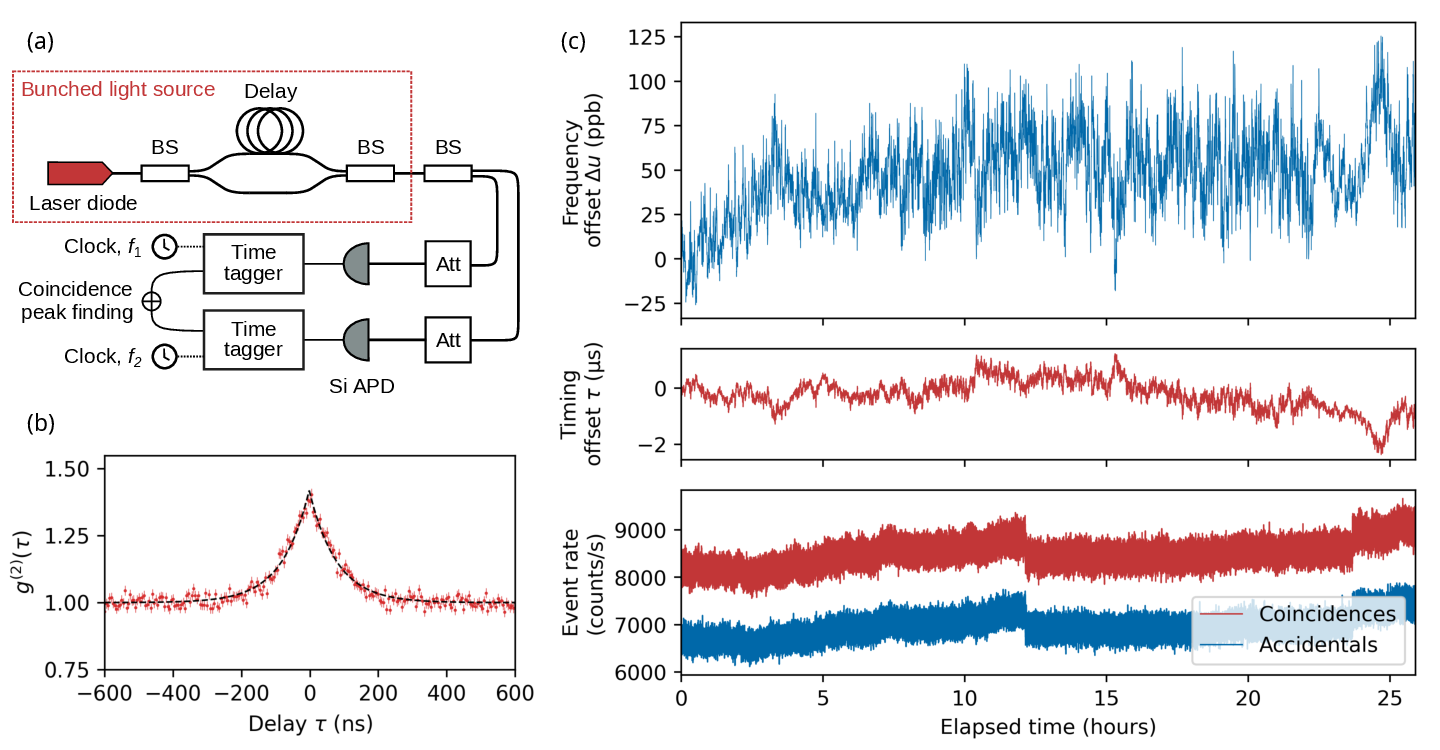}
  \caption{\label{fig:result:main}
    \textbf{(a)} Simplified experimental setup for clock
    synchronization. Bunched light is sent to two different parties through
    separate -102\,dB channels, before detection by Silicon avalanche
    photodetectors (Si APDs) and timestamping by time taggers referenced to independent crystal oscillators of frequencies $f_1$ and $f_2$. Singles count rate $s_1$ and $s_2$ of approximately 200\,kcounts/s are recorded on each side. BS: beamsplitter; Att: attenuators.
    \textbf{(b)} Second-order coherence function $g^{(2)}(\tau)$ of the bunched light source, with the curve fit in black corresponding to $g^{(2)}(0)=1.42(1)$ peak and coherence time $\tau_c=180(6)$\,ns. Error bars correspond to Poissonian errors from counting statistics.
    \textbf{(c)} Long-term frequency offset ($1 + \Delta{}u = f_1/f_2$), timing offset $\tau$, and event rates during a $25$\,hour clock synchronization run between independent time taggers running on separate quartz crystal oscillators, after an initial frequency correction of $4.0$\,ppm. Correlation peak tracking is performed using a $256$\,ns coincidence window. The sharp spike and dip in event rates are attributed to laser mode hopping.
  }
\end{figure*}

The experiment is shown Fig.\,\ref{fig:result:main}(a).
Our source of bunched light is formed by laser light of 780\,nm wavelength sent into an unbalanced Mach-Zehnder interferometer with optical delay longer than the laser coherence time $\tau_c$~\cite{yeoNearlosslessMethodGenerating2026} (see Appendix~\ref{appendix:setup:detailed} for a more detailed explanation).
Phase fluctuations in the laser result in an intensity correlation at the output, that can be measured through the second-order coherence function $\text{g}^{(2)}(\tau)$, given by
\begin{align}
  \text{g}^{(2)}(\tau)=1+\frac{1}{2}\exp\left(-\frac{2|\tau|}{\tau_{\text{c}}}\right).
  \label{eqn:g2source}
\end{align}
The brightness of the source is 1.55(2)\,mW, with a measured coherence peak of $g^{(2)}(\tau=0)=1.42(1)$ and coherence time of $\tau_c = 180(6)$\,ns, as shown in Fig.\,\ref{fig:result:main}(b).

The light from the source is shared between two parties with optical fiber
channel transmissions of $-101.9(4)$\,dB and $-102.2(4)$\,dB, respectively, using optical attenuators.
Photon arrival events are detected using Silicon avalanche photodetectors (APDs) on each side, with count rates of 192\,kcounts/s and 182\,kcounts/s without correcting for dark counts and afterpulsing.

The detection events are read by independent time taggers disciplined by different free-running 10\,MHz crystal oscillators.
The generated timestamps are then exchanged for coincidence peak finding and tracking in real-time.
The peak tracking program continuously serves a timing offset between each party by pairing photon detection events between both parties within a $256$\,ns coincidence window, over the span of the 25\,hour measurement.
The average coincidence rate of 8500\,events/s is consistently higher than the accidental count rate of 7000\,events/s, indicative of successful frequency tracking, as shown in Fig.\,\ref{fig:result:main}(c).

Crystal clocks drift in frequency due to temperature fluctuations and
electronic noise. We are able to reconstruct the resulting frequency offset $\Delta u'$ by
monitoring the drift in served timings using individual samples spanning
10.74\,s, with a resolution of 0.537\,s (see Fig.\,\ref{fig:result:main}(c)).

\begin{figure}
  \centering
    \includegraphics[width=\linewidth]{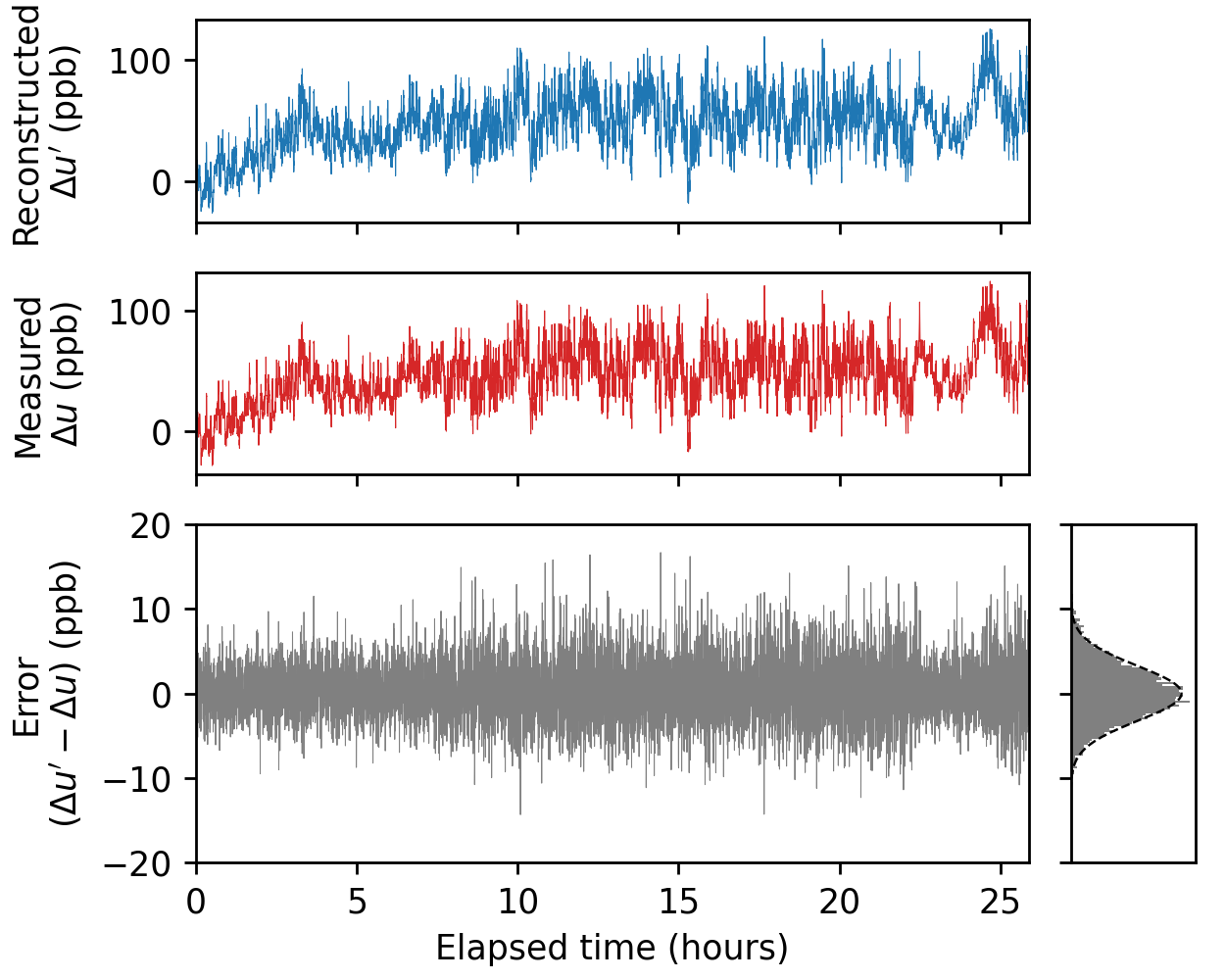}
  \caption{
    \label{result:freq:accuracy}
  A concurrent measurement of the actual frequency offset $\Delta u$ between the two external $10$\,MHz clocks, as well as the corresponding error in the reconstructed frequency offset $\Delta u'$. The accuracy of the frequency served by the peak tracking algorithm is $3.2$\,ppb on average.
  }
\end{figure}

The actual frequency offset $\Delta u$ was obtained via a heterodyne measurement using identical copies of the clock signals with integration time of 10\,s, shown in Fig.\,\ref{result:freq:accuracy}.
A maximum instantaneous frequency difference of 35.4\,ppb, corresponding to an average drift of 3.3\,ppb/s, was observed.

We estimate the error in the reconstructed signal by performing linear interpolation on the measurement and subsequent differencing.
The reconstructed frequency offset is found to be in good agreement with the measured frequency offset, with a root-mean-squared error of 3.2\,ppb averaged over the 25\,hour span.

\section{Peak finding}

Clock synchronization can be decomposed into two parts: first identifying the initial clock frequency offset and time difference between two parties, then monitoring the timing drift.
The former is performed by distributing photons with time-correlated statistics to each party, and finding the coincidence peak using cross-correlation to identify the time delay.

Efficient peak identification relies on the circular convolution theorem to
compute the cross-correlation between two sets of reconstructed detection time
traces $a[k]$ and $b[k]$,
\begin{equation}
  \label{eq:convolution}
    g^{(2)}(\tau) \sim(a \star b)[k] = \mathcal{F}^{-1}\!\left\{
      \overline{\mathcal{F}\{a\}} \cdot{} \mathcal{F}\{b\}
    \right\}\![k],
\end{equation}
under the discretization $\tau = k \delta{}t$ with $k \in \mathbb{Z}$ and time resolution $\delta{}t$, using the Fourier Transform $\mathcal{F}$ and its inverse\footnote{
  Algorithms faster than FFT's $\mathcal{O}(n \log{n})$ time complexity are also possible~\cite{calderaroFastSimpleQubitBased2020a,krauseClockOffsetRecovery2024}, but require additional encoding of time-synchronization strings within the light pulses.
}~\cite{ho2009clock}.
The success of peak identification is constrained by singles rate, rate of true coincidences, and choice of optimal parameters for the Fast Fourier Transform (FFT) to compute the time delay: specifically, the number of time bins $N = 2^q$ ($q\in\mathbb{Z}^+$), and initial bin width $\delta{}t$.

We model the peak finding probability --- in both cases, with and without a
reliable frequency reference between both parties --- and perform an
exhaustive parameter scan to identify appropriate FFT parameters.

\begin{figure}
  \centering
    \includegraphics[width=\linewidth]{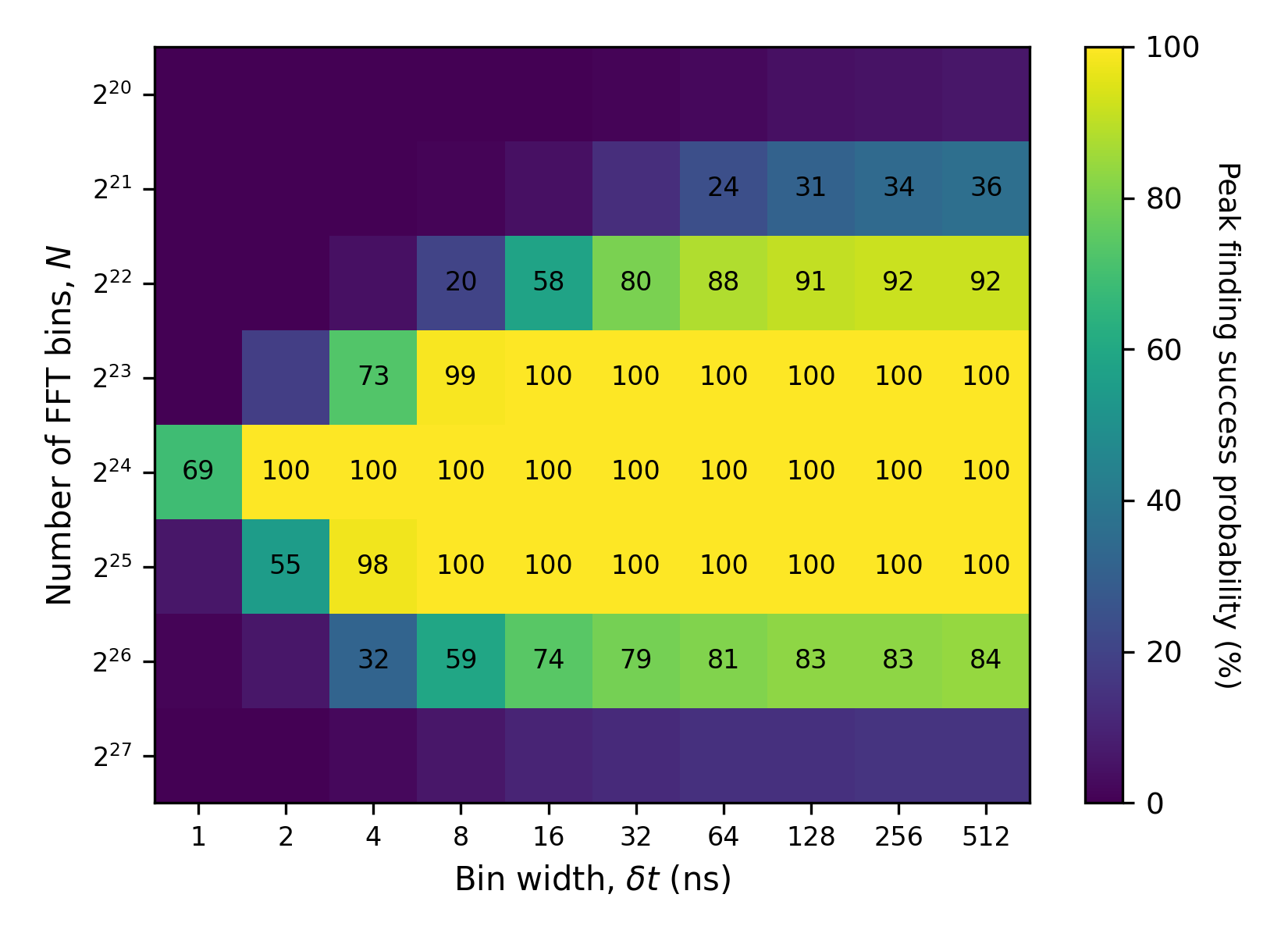}
  \caption{
    \label{fig:finding:surfaceplot}
    Success probabilities of finding the correct peak position, numerically obtained by performing $>\!\!10^7$ Monte Carlo trials for each $(N,\delta t)$ pair to estimate and interpolate the cumulative distribution in Eqn.\,\ref{eqn:peakfinding}.
    The parameters are singles detection rate $s_1=s_2=100$\,kcounts/s, coincidence rate $c=650$\,counts/s, bin overlap $\nu=0.5$, and frequency offset error of $\Delta{}u=50$\,ppb (precompensation step size of 100\,ppb), corresponding to the setup in Fig.\,\ref{fig:result:main} with an additional $3$\,dB attenuation per channel.
  }
\end{figure}

\subsection{No frequency offset}

The minimum acquisition time required for the cross-correlation to obtain a
flat cross-correlation noise floor is $T = N\delta{}t$.
This noise floor arises from accidental coincidences --- attributed to noise sources, such as coincidences with uncorrelated detection events and dark counts --- with an expected value of $C_a = (s_1 s_2 \delta{}t) T$, where $s_1$ and $s_2$ are the detection event rates on each channel.
The detection events are well-approximated by a Poisson distribution after binning, so the observed accidental coincidences $X_k$ in each time bin $k$ also follows a Poisson distribution $\mathcal{P}$ with mean $\lambda = C_a$, i.e., $X_{k\in\{1,\ldots{},N\}} \sim \mathcal{P}(X_k=x_k|\lambda=C_a)$.

The maximum observed value across all (identically distributed) time bins $X_{(N)} \equiv \max\{X_k\}$ is therefore given by the max-order Poisson distribution (see Appendix~\ref{appendix:derivation}) whose probability distribution is
\begin{equation*}
  f_{X_{(N)}}(x) = \left[F_X(x|\lambda{})\right]^N - \left[F_X(x|\lambda{}) - f_X(x|\lambda{})\right]^N,
\end{equation*}
where $f_X$ and $F_X$ correspond to the probability mass and cumulative distribution functions of a single bin.

The coincidence rate above background accidentals in a single time bin (denoted $c_e$) required to be identified as the highest peak in the cross-correlation is thus
\begin{equation}
c_e > \frac{1}{T}\left(X_{(N)} - X\right).
\label{eqn:coincnofreq}
\end{equation}

The coincidence rate per time bin can be maximized by setting the timing resolution $\delta{}t$ to be of the same scale as the coherence time $\tau_c$, so that most of the coincidence events fall within the same time bin, i.e., $\delta{}t \sim \tau_c$.
Some of these events may fall into an adjacent time bin instead due to off-centered bins (since the exact time offset is not known \textit{a priori}): this introduces an adjustment factor into Eqn.\,\ref{eqn:coincnofreq} representing the degree of bin overlap $\nu \in [0.5, 1]$, yielding
\begin{equation}
c_e > \frac{1}{\nu{}T}\left(X_{(N)} - X\right).
\label{eqn:coincnofreq:binoverlap}
\end{equation}

\subsection{Variable frequency offset}

In the case of two separate clocks with slightly different clock frequencies, the timing delay between photon arrivals between each party can drift by $\Delta{}\tau_{\text{drift}} \approx T\Delta{}u(t)$ due to the non-zero clock-frequency offset $\Delta{}u(t)$ after elapsed time $T$.
For a sufficiently small $T$, the frequency offset can be approximated as a constant, i.e., $\Delta{}u(t) = \Delta{}u$.

In order to maximize the number of coincidences in a single time bin, the time bin should ideally be as wide as the timing drift, i.e., $\delta{}t \geq \Delta{}\tau_\text{drift}$, or in other words, $N \leq 1/\Delta{}u$. We model this as an additional correction factor $\xi \equiv \max{\{1, N\Delta{}u\}}$, and together with Eqn.\,\ref{eqn:coincnofreq:binoverlap} yields the minimum required true coincidence rate for successful peak finding given by
\begin{equation}
c_e > \frac{\xi}{\nu{}T}\left(X_{(N)} - X\right).
\label{eqn:peakfinding}
\end{equation}

The surface of Eqn.\,\ref{eqn:peakfinding} is shown in Fig.\,\ref{fig:finding:surfaceplot}, by performing Monte Carlo simulations across different $N$ and $\delta{}t$. Our model provides better estimations of the peak finding probability compared to previous works~\cite{ho2009clock,spiessClockSynchronizationCorrelated2023} by avoiding the normal approximation to the noise in the FFT (see Appendix~\ref{appendix:poissonprobability}).

The reduction in coincidence counts due to the presence of frequency offset can be mitigated either by choosing a smaller $N$, or by performing a frequency precompensation sweep~\cite{spiessClockSynchronizationCorrelated2023} on the set of timestamps $\{t_i\}$ to reduce the apparent $\Delta{}u$ between the two clocks and allow for larger $N$ values. The compensation is given by the mapping
\begin{align}
  t_i \,&\rightarrow{}\, \Delta{}t_i ( 1 + \Delta{}u) + t_{i-1} = t_i + \Delta{}t_i \Delta{}u,
  \label{eqn:timingcorrection}
\end{align}
where $\Delta{}t_i \equiv t_i - t_{i-1}$ is the separation between consecutive timestamps.

In practice, while the frequency offset between crystal clocks can be high ($\sim{}\!10$\,ppm), the short-term stability of the clocks themselves is much better ($<\!10$\,ppb/s).
This means the full precompensation sweep need not be repeated even if clock synchronization were lost; using the last estimated frequency offset $\Delta u_0$ as well as $\Delta u_0\pm100\,\text{ppb}$ is typically sufficient for peak recovery.

Once an initial peak has been found, the frequency and timing resolution can be further improved by repeating the respective corrections with progressively smaller time bins, until the desired resolution has been reached~\cite{ho2009clock}.
Our optimized implementations of peak finding and frequency compensation are open-sourced~\cite{fpfind}.

\section{Peak tracking}

\begin{figure}
  \includegraphics[width=\linewidth]{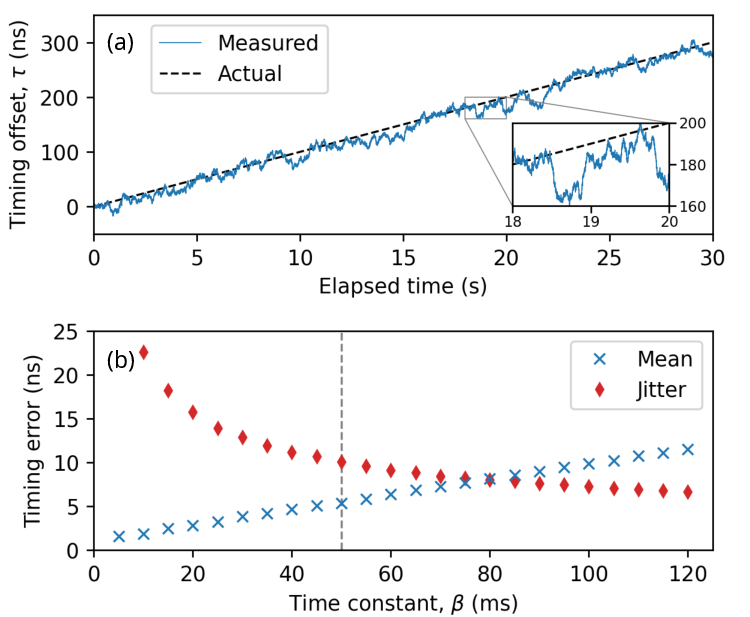}
  \caption{\label{fig:tracking}
    Characterization of peak tracking accuracy under a constant frequency offset $\Delta{}u=10$\,ppb, with a time constant $\beta=50$\,ms chosen for the exponential moving average filter used in the experiment.
    \textbf{(a)} Measurement of reported timing offsets (blue) against the actual offset (black) over a period of 30\,seconds. The inset highlights a $\approx{}\!30$\,ns timing error over a 0.5-second window, which suggests that a histogram fit does not improve accuracy.
    \textbf{(b)} Measurements of average offset error and jitter across different time constants $\beta$, over a measurement period of 10\,minutes. Timing jitter is equivalent to the root-mean-squared error, and corresponds to 10\,ns with $\beta=50$\,ms (marked in gray dashed line). Peak tracking fails with time constants less than 5\,ms.
  }
\end{figure}

The use of crystal oscillators as frequency references results in changes in
frequency offset over time, due to their long-term stability in $\Delta{}u(t)$
of up to 100\,ppb/day.
This causes a drift in the correlation peak position over time which needs to be tracked.

Tracking can be performed by looking for coincidences within a sufficiently wide coincidence window, such that drifts in the peak are captured.
However, when using bunched light for peak tracking, coincidences events are effectively dominated by background accidentals due to the low $g^{(2)}(\tau)$ of the light source.
Directly returning each time difference found will result in the peak being quickly lost.

We apply a smoothing operation using an exponential moving average filter to introduce damping against events far from the current estimated peak position.
This effectively minimizes noise fluctuations and allows for the peak to be tracked.

In order to quantify the accuracy of the peak tracking, we supply a constant $10$\,ppb frequency offset between the two timestamps using a function generator and derive the true timing offset.
Using a time constant $\beta=50$\,ms for the moving average filter, a timing jitter of 10\,ns can be obtained, as shown in Fig.\,\ref{fig:tracking}, which is an order of magnitude better than the $\tau_c=180(6)$\,ns coherence time of the source.
The jitter was tested to remain the same even under larger frequency offsets of 50\,ppb as well.

A tracking lag is also present and can be seen in Fig.\,\ref{fig:tracking}, where the measurement best-fit line is slightly displaced from the true timing offset. This is because the frequency offset is non-zero and constant, and manifests as a mean timing error of around $6$\,ns at $\beta=50$\,ms.
In the case of free-running clocks after frequency compensation, this lag is expected to be insignificant due to the frequency offset fluctuating about zero.

We also make a small note regarding the use of timing offset histograms to improve timing resolution.
For photon pair and pulsed sources, higher timing resolution can be achieved by collecting a histogram within a small timing window and extracting the peak from a normal distribution curve fit, due to the narrow peak coherence signature of $\tau_c\ll{}1$\,ns~\cite{spiessClockSynchronizationCorrelated2023}.
However, for our bunched light source with a significantly noisier coherence peak signature, this technique does not improve the timing accuracy, as can be seen in the served timing offset in the inset of Fig.\,\ref{fig:tracking} being consistently far from the mean.

Active frequency compensation is also performed online to keep the frequency offset small, to avoid peak tracking loss since the correlation peak is less likely to drift out of the coincidence window.
We do this by estimating the frequency offset $\Delta{}u$ from the rate of change in timing drift (see Appendix~\ref{appendix:math:tracking}), then performing timing compensation for each timestamp using the estimated frequency offset, identical to that of Eqn.\,\ref{eqn:timingcorrection}.

\section{Conclusion}

We demonstrated successful clock synchronization between two parties over a symmetrical -102\,dB transmission channel, using a bunched light source of coherence time $\tau_c = 180(6)$\,ns. Peak tracking was performed online over a span of 25\,hours with active frequency compensation and an exponential moving average filter of time constant $\beta=50$\,ns, achieving an overall timing jitter of 10\,ns.

Previous papers on clock synchronization performed using photon pair sources rely on a peak significance metric for peak finding~\cite{ho2009clock,spiessClockSynchronizationCorrelated2023}, used as a threshold for quantifying the probability of the cross-correlation peak being attributed to noise.
While it remains a useful metric for estimating peak location, this underestimates the peak finding probabilities under low signal conditions with small time bins due to the use of normal approximation.
We develop a model that accounts for the Poisson nature of the coincidences, and calculate instead the true probability of the cross-correlation peak being the signal, as well as derive optimal FFT parameters for a given true coincidence and singles rate.
This remains applicable even when using any other sources of timing correlations, including photon pairs and thermal light.

The clock synchronization scheme in this work can be directly applied to protocols that use photon bunching as a resource, or indirectly by software timing compensation using the served timing and frequency offsets.
Direct frequency compensation can also be performed by actively tuning the crystal oscillator frequencies, so that the clock signals themselves can be utilized as part of a clock distribution network.

We can also take advantage of the fact that the second-order coherence is preserved across arbitrary partitioning of the light, to distribute the signal amongst multiple parties in a star topology.
Since splitting light into two separate channels incurs an additional insertion loss of $3$\,dB per channel, clock distribution to $2^n$ parties can be achieved with only $-3n$\,dB of additional loss per channel, e.g., a 128-party setup with this source incurs about $-80$\,dB.

This work paves the way towards clock synchronization using telecommunication O-band and C-band bunched light, which will be able to propagate with minimal chromatic dispersion over longer distances, e.g., 500\,km of G.652 telecommunication fiber with 0.2\,dB/km optical loss at 1550\,nm.
This additionally opens up the possibility of using erbium-doped fiber amplifiers to amplify the correlation signal.
While timing offset changes as a result of thermal expansion in optical fibers will need to be addressed, this effect is fundamentally indistinguishable from a clock frequency drift in this scheme and can hence be compensated for in the same manner.

\section{Acknowledgements}
The authors would like to thank Ng Boon Long for helpful discussions in
setting up the concurrent frequency measurement.
This research is supported by the National Research Foundation, Singapore
through the National Quantum Office, hosted in A*STAR, under its Centre for
Quantum Technologies Funding Initiative (S24Q2d0009) and the Quantum
Engineering Programme 3.0 under grant W25Q3D0004.

\bibliography{bibliography.bib}

\appendix
\input{appendix}

\end{document}

%% file: appendix.tex
\section{Bunched light source and experimental setup}
\label{appendix:setup:detailed}

The experimental setup in this work uses light from a distributed-feedback (DFB) laser diode of wavelength 780\,nm, coupled into 780-HP fiber to project into a single optical mode. This light is sent into an unbalanced fiber-based Mach-Zehnder interferometer with 400\,m delay fiber. This corresponds to a $\Delta=2\,\mu\text{s}$ delay, which is longer than the coherence time $\tau_{\text{c}}=180\,\text{ns}$.

Assuming that the power and polarization in both arms match, the output field from one of the interferometer ports is given by
\begin{equation*}
	E(t)=\frac{1}{2}[E_0(t)+E_0(t+\Delta)],
\end{equation*}
where $E_0(t)$ is the electric field of the laser light. The overall second-order coherence function thus gives
\begin{align*}
g^{(2)}(\tau) = &\>\frac{1}{4}\left[g_0^{(2)}(\tau+\Delta) + g_0^{(2)}(\tau-\Delta)\right]\\
&+\frac{1}{2}\left[g_0^{(2)}(\tau) + \vert g_0^{(1)}(\tau)\vert^2\right]\\
= &\>1 + \frac{1}{2}\exp\left(-\frac{2|\tau|}{\tau_c}\right),
\end{align*}
where $g_0^{(2)}(\tau)=1$ is the second-order coherence function of the laser light and $g_0(\tau)=\exp\left(-2|\tau|/\tau_c\right)$ is the first-order coherence function of the laser light, assuming a Lorentzian spectral profile. The peak at $g^{(2)}(\tau=0)$ is a manifestation of the photon bunching effect that saturates at 1.5 in our setup. Our measured peak value of $1.42(1)$ is attributed to the fact that the laser is not fully coherent~\cite{yeo2024crossg2} and other experimental imperfections. A more in-depth explanation and discussion about the source can be found in \cite{yeoNearlosslessMethodGenerating2026}.

A second BS is used to further split the light into two symmetrical channels
for downstream detection using fiber-pigtailed active-quenched Si avalanche
photodiodes (Excelitas SPCM-800-10-FC). Attenuation in each channel is
achieved by cascading fiber beamsplitters as well as fiber decoupling at the
mating sleeves, and is measured to be relatively stable over multiple days.

Time tagging is subsequently performed by timestamp devices (S-Fifteen
Instruments TDC2) with a 1-$\sigma$ timing jitter of 20\,ps.
The clock to each timestamp device is supplied by external 10\,MHz crystal oscillators without any temperature stabilization.

The actual frequency offset between the two clocks was measured concurrently with the clock synchronization experiment. The $10$\,MHz signal was duplicated and combined in a frequency mixer (Mini-Circuits ZFM-2+) followed by a low-pass filter (Mini-Circuits SLP-2.5+, $2.5$\,MHz cut-off). The mixed signal was then sampled by an oscilloscope at a rate of $2.5$\,ksamples/s over $10$\,s, which performed a $2^{14}$-bin FFT (nominal resolution of approximately $0.15$\,Hz) with a Hann window.

The frequency offset error is then calculated by measuring the difference between the served frequency offset (from the frequency estimation step in the clock synchronization) and the measured frequency offset with linear interpolation. The histogram of offset errors is fitted using a Gaussian probability density function, obtaining a standard deviation of $3.2$\,ppb as seen in Fig.\,\ref{result:freq:accuracy}. 

\section{Derivation of max-order distribution}
\label{appendix:derivation}

The FFT for peak finding according to Eqn.\,\ref{eq:convolution} is performed with $N$ bins and accidental coincidences in each of these bins follow the Poisson distribution $\mathcal{P}$ with mean $\lambda{}$. The maximum value across all bins increases with the number of bins. The distribution of the max-order statistic (i.e., the maximum of all bins) with respect to the number of coincidences $x$ and number of bins $N$ of mean value $\lambda{}$ is denoted $f_{(N)}(x|\lambda)$.

Since each bin is assumed to be independent and identically distributed, the corresponding max-order cumulative distribution function (CDF) can be expressed as the product of the CDF of individual bins,
\begin{align*}
F_{(N)}(x|\lambda{}) = \underbrace{F(x|\lambda{})\times\ldots{}\times{}F(x|\lambda{})}_{N\ \text{times}} = \left[F(x|\lambda{})\right]^N.
\end{align*}

We therefore express the max-order probability mass function (PMF) in terms of the Poisson PMF and CDF of a single bin,
\begin{align*}
  f_{(N)}(x|\lambda)
  &= F_{(N)}(x|\lambda{}) - F_{(N)}(x-1|\lambda{}) \\
  &= \left[F(x|\lambda{})\right]^N - \left[F(x-1|\lambda{})\right]^N \\
  &= \left[F(x|\lambda{})\right]^N - \left[F(x|\lambda{}) - f(x|\lambda{})\right]^N.
\end{align*}

Notably, this form is already tenable for direct computation without further simplification, even though computing the difference of $n^{\text{th}}$-powers generally causes catastrophic cancellation due to floating-point rounding errors.
The max-order Poisson PMF distribution width heuristically scales roughly with $\sqrt{\lambda{}}$, which requires the PMF to be accurate to at least $1/\sqrt{\lambda{}}$, e.g., $\lambda{}\le 10^{4}$ needs at least $\sim{}10^{-2}$ accuracy.
Since $F(x|\lambda{}) \in [0,1]$ and the fact that exponentiation can be easily applied for large $N$ using numerical techniques (such as exponentiation-by-squaring), the rounding errors can be minimized to near the floating-point precision, e.g., $\sim{}10^{-16}$ for 64-bit floats.

For larger $\lambda{} > 10^{4}$ (i.e., high accidental coincidence rates per time bin), the normal approximation for each bin remains appropriate,
\begin{align}
f(x) \sim{} \mathcal{P}(x|\lambda{}) \approx{} \mathcal{N}(x|\mu=\lambda{},\sigma=\sqrt{\lambda{}}),
\label{eq:normalapprox}
\end{align}
and the corresponding max-order probability distribution function follows a more tractable form for numerical computation,
\begin{align*}
  f_{(N)}(x|\mu,\sigma)
  &= \frac{d}{dx} F_{(N)}(x|\mu,\sigma) \\
  &= \frac{d}{dx} \left[F(x|\mu,\sigma)\right]^N \\
  &= N f(x|\mu,\sigma) \left[F(x|\mu,\sigma)\right]^{N-1}.
\end{align*}
This in fact corresponds to the first-order term for the discrete case after a binomial expansion,
\begin{align*}
  f_{(N)}&(x|\lambda) \\
  &= \left[F(x|\lambda{})\right]^N - \left[F(x|\lambda{}) - f(x|\lambda{})\right]^N \\
  &= \left[F(x|\lambda{})\right]^N - \left[\left[F(x|\lambda{})\right]^N \right. \\
  &\quad \left. - N f(x|\lambda{})\left[F(x|\lambda{})\right]^{N-1} + \ldots\right] \\
  &= N f(x|\lambda{})\left[F(x|\lambda{})\right]^{N-1} - \mathcal{O}\left(f^2F^{N-2}\right),
\end{align*}
noting that $f(x|\lambda)/F(x|\lambda) \ll 1$ at large $\lambda$.

\section{Comparison of peak finding probabilities with Poisson / normal approximation}
\label{appendix:poissonprobability}

\begin{figure*}
  \includegraphics[width=\textwidth]{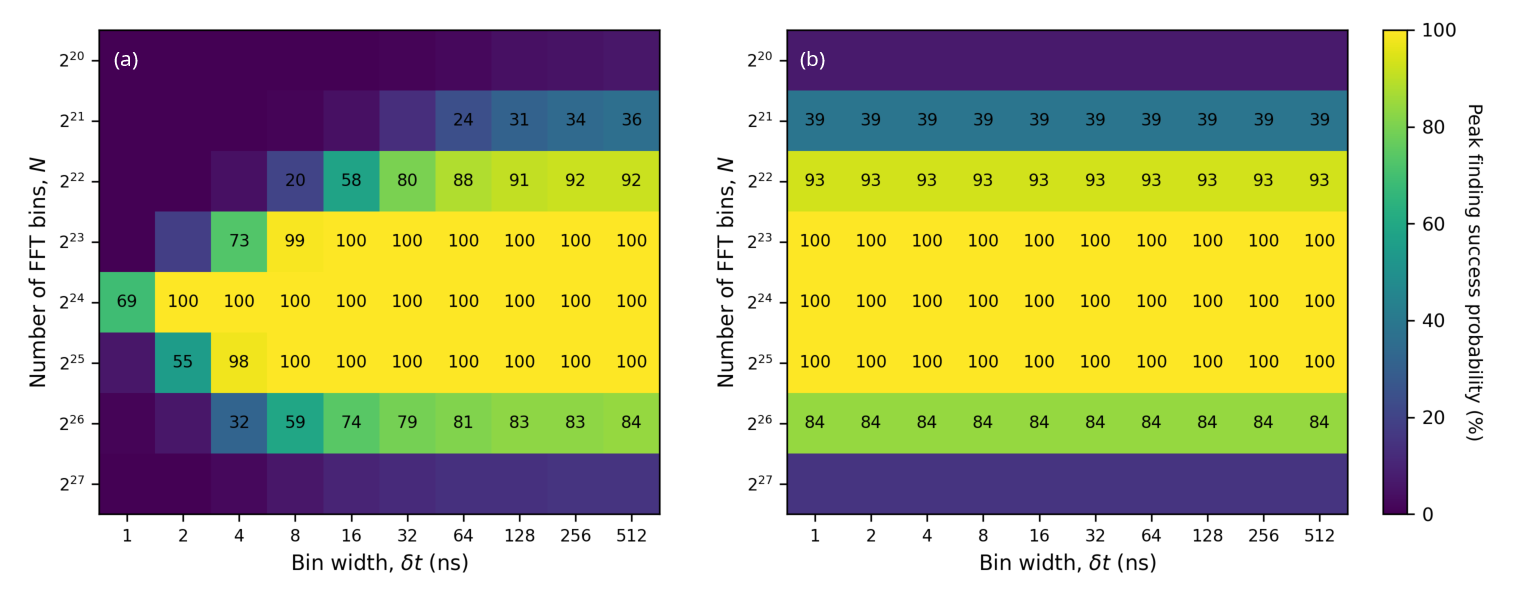}
  \caption{\label{finding:gaussian}
    \textbf{(a)} Success probabilities of finding the correct peak position by solving Eqn.\,\ref{eqn:peakfinding}, given singles detection rate $s_1=s_2=100$\,kcounts/s, coincidence rate $c=650$\,counts/s, and bin overlap of $\nu=0.5$, and frequency offset error of $\Delta{}u=50$\,ppb (precompensation step size of 100\,ppb).
    \textbf{(b)} Corresponding probabilities using the normal approximation for observed accidental coincidences $X$ in Eqn.\,\ref{eqn:peakfinding}, with the same parameters. The probabilities are strongly dependent on $N$ and minimally with $\delta t$, similar to the estimations provided using the statistical significance framework.
  }
\end{figure*}

To understand the effect of normal approximation on peak finding success probabilities, it is useful to first understand the use of the max-order distribution.

Previous models derive a statistical significance $S$ representing the number of standard deviations $\sigma_X$ the coincidence peak value deviates from the mean number of accidental coincidences $X$ (of a \textit{single} time bin).
Ignoring any corrections for frequency difference, the significance shares the form
\begin{align}
S = \frac{c_e T}{\sigma_{X}} = \frac{c_e T}{\sqrt{s_1s_2 T \delta t}} = c_e \sqrt{\frac{N}{s_1 s_2}}
\label{eqn:significance}
\end{align}
using notation consistent with this paper (corresponding to Eqn.\,8 of \cite{ho2009clock} and Eqn.\,A3 of \cite{spiessClockSynchronizationCorrelated2023}). Notably, this equation only depends on the number of time bins $N$ and has no dependence on time bin width $\delta t$.

The search for the true coincidence peak is more concerned with distinguishing the true coincidence peak from the \textit{next highest} peak arising from noise, and less to do with the average fluctuations of a single time bin.
Since the number of accidental coincidences $X$ in each time bin is stochastic, the maximum observed noise peak across all time bins intuitively increases with increasing number of observations (i.e., number of bins $N$).
This maximum is distributed according to the order statistics $X_{(N)}$ of all time bins.

In contrast, the derivation in Eqn.\,\ref{eqn:significance} implicitly assumes that the noise peak is given only by the accidental coincidences $X$ from a single time bin. The use of statistical significance is thus more accurately described as an indication of the \textit{peak misidentification rate}: the probability that the true coincidence peak is misidentified as noise, given that the location of said peak is \textit{already known}. Our derivation of Eqn.\,\ref{eqn:peakfinding} in the main text, which accounts for $X_{(N)}$, yields the actual peak identification rate.

Due to the positive excess kurtosis (i.e., ``fatter tail'') of the Poisson distribution relative to the normal distribution, the corresponding max-order distribution $f_{(N)}(x|\lambda)$ is skewed more towards higher peak values.
At low coincidence rates per time bin ($\lambda \lesssim 10^4$) occurring at small $N$ and $\delta t$ values, the normal approximation $f_{(N)}(x|\mu,\sigma)$ underestimates the noise peak value, and thus overestimates the peak finding success probabilities.

The success probabilities for the Poisson model and the normal approximation model are shown in Fig.\,\ref{finding:gaussian} by numerically solving Eqn.\,\ref{eqn:peakfinding} using Monte Carlo simulations of $>10^7$ trials per $(N,\delta t)$ pair.
The normal approximation indeed overestimates the peak finding probabilities at small time bin width $\delta t$. We use $N = 2^q$ ($q\in\mathbb{Z}^+$) for its logarithmic scale.

\vspace{-1em}
\section{Derivation of peak tracking equation and parameters}
\label{appendix:math:tracking}

We perform active frequency compensation by estimating the clock frequency offset from the set of timestamps.
The timing difference $\tau_i := t_i - t'_i$, between a timestamp pair $\{t_i,t'_i\}$ from both parties, is continuously served by searching for photon pair detection events within a prescribed coincidence tracking window.

The frequency offset $\Delta{}u$ is given by the ratio between measured elapsed time $\Delta{}t_i := t_i - t_{i-1}$ with respect to some reference elapsed time, in this case the elapsed time measured by the other peer $\Delta{}t'_i$,

\begin{align*}
\frac{\Delta{}t_i}{\Delta{}t'_i} = 1 + \Delta{}u.
\end{align*}
Rewriting in terms of the measured successive timing difference, we can estimate the frequency offset by measuring the rate of change in the timing difference, i.e.,
\begin{align*}
\Delta{}u_i = \frac{\tau_i - \tau_{i-1}}{\Delta{}t'_i}.
\end{align*}

Active frequency compensation is therefore achieved by performing a timing correction for each timestamp $t_i$ using the estimated frequency offset,
\begin{align*}
t_i &\rightarrow{} \Delta{}t_i ( 1 + \Delta{}u) + t_{i-1} = t_i + \Delta{}t_i \Delta{}u,
\end{align*}
where the overall frequency offset $\Delta{}u$ accumulated is also given by
$\Delta{}u = \prod_0^i{\left(1 + \Delta{}u_i\right)} - 1$.

Peak tracking with thermal light is intrinsically noisy due to the signal being dominated by accidental coincidences.
Hence, we use an exponential moving average filter given by,
\begin{align*}
\tau{}'_i = \alpha\tau_{i} + (1 - \alpha)\tau'_{i-1},\quad{} \tau'_0 = \tau_0,
\end{align*}
which behaves like a low-pass filter to smooth the signal. The coefficient $\alpha$ represents the relative weight of timing offset $\tau_i$ with respect to the accumulated average $\tau'_{i-1}$, and can be adjusted depending on the observed signal rate.

Given a unit step impulse, the time it takes to reach $1 - 1/e$ of the signal is associated with a time constant $\beta$ related to the coefficient $\alpha$,
\begin{align*}
\alpha = 1 - \exp{\left(-\frac{\overline{\Delta{}t}}{\beta}\right)} \approx{} \frac{\overline{\Delta{}t}}{\beta},\qquad{} \beta \gg \overline{\Delta{}t},
\end{align*}
after Taylor expansion of the exponential, where $\overline{\Delta{}t}$ is the average separation time between consecutive timestamp events. The singles rate $s$ in our experiment is 200\,kcounts/s, so $\overline{\Delta{}t} = 1/s = 5$\,\textmu{}s. We set the time constant $\beta = 50$\,ms, which is 4 orders of magnitude longer than $\overline{\Delta{}t}$ for averaging.